\documentclass[aps,preprint,nofootinbib]{revtex4}
\usepackage{graphicx,epstopdf,color}

\begin{document}

\title{Double-action dark matter, PAMELA and ATIC}

\author{Kingman Cheung$^{1,2,3}$,
Po-Yan Tseng$^1$, and Tzu-Chiang Yuan$^4$
}

\affiliation{$^1$Department of Physics, National Tsing Hua University, 
Hsinchu 300
\\
$^2$Physics Division, National Center for Theoretical Sciences,
Hsinchu 300
\\
$^3$Division of Quantum Phases \& Devices, School of Physics, 
Konkuk university, Seoul 143-701, Korea
\\
$^4$Institute of Physics, Academia Sinica, Nankang, Taipei 11529, Taiwan
}

\date{\today}

\begin{abstract}
Motivated by a two-bump (or 1-peak plus 1-hump) structure in the ATIC data,
we perform a statistical analysis fitting the PAMELA and ATIC data
to a dark matter model, in which the dark matter particle can undergo
both annihilation and decay.   Using a chi-square analysis we show that
both data can be simultaneously fitted better with such a double-action dark 
matter particle.  
We use an existing neutrino mass model in literature to
illustrate the idea.
\end{abstract}

\maketitle

\section{Introduction}

The year 2008 had been filled with excitement from a number of dark
matter (DM) experiments.  The PAMELA Collaboration \cite{pamela-e} has
reported an unexpected rise of positron fraction at the energy range 
of $10-100$ GeV, unlike the power-law falling background.  
This provides further support to the earlier results reported from
HEAT \cite{heat} and AMS-1 \cite{ams-01}.
However, similar enhancement of the anti-proton flux was expected but not seen 
by PAMELA \cite{pamela-p} provides a challenging puzzle.
Other surprises came from two balloon experiments ATIC \cite{atic}  
and PPB-BETS \cite{ppb-bets} at the
South Pole Antarctica. 
The ATIC data showed an excess of galactic cosmic-ray electrons/positrons at
energies of $300 - 800$ GeV. 
These experimental results have stimulated a lot of
theoretical speculations about possible mechanisms, including dark matter 
annihilation
\cite{dm-anni}, decaying dark matter with a very long lifetime \cite{dm-dec}, 
or simply astrophysical origins from either 
ultrahigh energy cosmic rays \cite{cosmic} 
or nearby pulsars \cite{pulsa} within a few kilo-parsec.
If the observed positron excess is indeed 
due to dark matter annihilation, the data sets require an
annihilation cross section 
$\langle \sigma v \rangle$
of the order of $10^{-23}\; {\rm cm}^3\;{\rm s}^{-1}$, 
which is two to three order of magnitudes larger than naively expected from a
thermo-WIMP dark matter in most popular models 
like the minimal SUSY and Kaluza-Klein models.  
Either a large boost factor or Sommerfeld-type
enhancement \cite{sommer} can be used to explain such a large
annihilation cross section.  On the other hand, a very long lifetime of the
decaying dark matter of the order of $10^{26}$ seconds is required to fit
the data.  Such a long-lived dark matter is consistent with other
cosmological constraints on our Universe.  Implications for further
investigations in future gamma-ray experiments and neutrino telescopes
have been studied \cite{impli}.

We point out that the excess in the ATIC data in fact consists of 
1-peak plus 1-hump structure.  The peak is from $300-800$ GeV while the bump
from $80-300$ GeV.  A possible explanation is that the dark matter
particle in the Universe undergoes both annihilation and decay.
The annihilation gives rise to the peak around 600 GeV while the
decay is responsible for the small bump around $80-300$ GeV.
The resulting positron fraction observed at PAMELA is a combination of
annihilation and decaying contributions.  

In this work, we perform a $\chi^2$ analysis which shows that the
fitting using a single mode (either annihilation or decay) is far 
less satisfactory than the double mode (both annihilation and decay).
The study shows that a dark matter particle of about 640 GeV with
a monochromatic annihilation spectrum and a soft decaying spectrum
is the best simultaneous fit to the PAMELA and ATIC data.  

The PAMELA data is highly restrictive on the anti-proton mode of the 
dark matter annihilation or decay \cite{pamela-p}.  
It points to the hint that the dark 
matter particle
may be lepto-philic or carrying a lepton number.  There are some 
models that propose the TeV right-handed neutrino, which is 
responsible for neutrino mass, to be the dark matter candidate.
We will borrow an example in literature \cite{Krauss,seto,Aoki} 
to illustrate the possibility. 

Some of the highlights in this paper include
\begin{itemize}
\item
  We show in a more quantitative way how the models are fitted 
to the ATIC and PAMELA data simultaneously;
\item
  We use the MINUIT program from the CERN library to vary the 
parameters of the model to minimize the
$\chi^2$ in order to obtain the best fit of the model;
\item
  We show that the dark matter candidate that can both annihilate and decay
can fit better to the data.
\end{itemize}

The organization of the paper is as follows. In the next section, 
we summarize the formulae we employed in our analysis for the
annihilation and decay of the dark matter particle 
without referring to any particular dark matter models. 
In Sec. III, we show our numerical analysis of $\chi^2$ fits.  
In Sec. IV, we describe a slight modification of 
an existing  dark matter model in literature 
that may give rise to the double-action.
We conclude in Sec. V.

\section{Double-Action Dark Matter}

\subsection{Dark Matter Annihilation}

Assuming a steady state condition while solving the 
diffusion equation for the positron as it traversed across 
the universe, its flux at Earth can be casted 
into the following semi-analytical form \cite{semianalytical-a,semianalytical-b}
\begin{equation}
 \Phi_{e^+} (E) = \frac{ v_{e^+} } { 4 \pi} \, f_{e^+} (E) \;,
 \label{semiflux} 
\end{equation} 
with $v_{e^+}$ close to the velocity of light $c$ and the function 
$f_{e^+} (E)$ is given by
\begin{eqnarray}
  f_{e^+} (E) &=& B \frac{1}{b(E)} \, \int_E^{E_{\rm max} }\; 
d E' \, I(\lambda_{D}(E,E') )\, Q_{\rm ann} (E') \nonumber \\
&=& 
B \frac{1}{b(E)} \, \eta \left( \frac{\rho_{\rm dm} }{M_{\rm dm}} \right )^2 
\, \sum \langle \sigma v \rangle_{e^+} \;
\int_E^{E_{\rm max} }\; 
d E' \, I(\lambda_{D}(E,E') )\,  \frac{d N_{e^+}}{ d E'_{e^+} } \;,
\label{semi}
\end{eqnarray} 
with $E_{\rm max} = M_{\rm dm}$ in the case of annihilating DM,
$M_{\rm dm}$ and $\rho_{\rm dm}$ are the mass and the density of the dark matter
respectively, and the overall constant $B$ is the boost factor.
In Eq.(\ref{semi}) we have expressed the source term 
$Q_{\rm ann}$ according to
\begin{equation}
Q_{\rm ann} = \eta \left( \frac{\rho_{\rm dm} }{M_{\rm dm}} \right )^2 
\, \sum \langle \sigma v \rangle_{e^+} \, \frac{d N_{e^+}}{ d E_{e^+} } \;,
\end{equation}
where $\eta = 1/2 (1/4)$ for Majorana or Dirac particle.
The summation is over all possible channels that can produce positron in 
the final state, and $dN_{e^+}/dE_{e^+}$ denotes the spectrum of the positron
energy per annihilation in that particular channel. We have suppressed
the index labeling the various channels to avoid notation cluttering 
in the equations.
The halo function $I(\lambda_D)$ can be parametrized by
\begin{equation}
  I(\lambda_D) = a_0 + a_1 \tanh \left( \frac{ b_1 - l}{c_1} \right )
\, \left[ a_2 \exp(- \frac{(l - b_2)^2}{c_2}) + a_3  \right  ]
\end{equation}
with $l = \log_{10} \left( \lambda_D/{\rm kpc} \right)$ and
the diffusion length $\lambda_D(E,E')$ is given by
\begin{equation}
 \lambda^2_D = 4 K_0 \tau_E \left[ \frac{ (E'/{\rm GeV} )^{\delta - 1}
                                        -(E/{\rm GeV} )^{\delta - 1} }{\delta-1}
 \right ] \;\; .
\end{equation}
The constants $a_{0,1,2,3}, b_{1,2}, c_{1,2}$ and $\delta$, $K_0$ can be found
respectively in Table 2 and Eq.(11) of Ref.~\cite{cirelli}. 
The energy loss rate function $b(E)$ in Eq.(\ref{semi}) is
\begin{equation} 
 b(E) = \frac{ E^2} { ({\rm GeV} \times \tau_E)}
\end{equation}
where $\tau_E = 10^{16}$ seconds.
In our analysis, we simply employ the monochromatic annihilation spectrum
\begin{equation}
  \frac{d N_{e^+} }{d E_{e^+}} (E) = \frac{1}{M_{\rm dm}} \delta 
\left ( 1 - \frac{E}{M_{\rm dm} } \right ) \;.
\end{equation}
Analogous formulas for the electron will be omitted here.
We only consider the monochromatic $e^-e^+$ spectrum in this study
for simplicity and clarity.  We could also add the muon and tau
channels in the annihilation, but that would introduce more parameters 
to complicate our analysis.

\subsection{Decaying Dark Matter}

The source term for a decaying dark matter in a particular channel is
\begin{equation}
Q_{\rm dec} = \frac{1}{\tau_{\rm dm} } \, 
 \left( \frac{\rho_{\rm dm} }{M_{\rm dm}} \right )
\,  \frac{d N_{e^+}}{ d E_{e^+} } \;,
\end{equation}
where $\tau_{\rm dm}$ is the lifetime of the DM and $dN_{e^+}/dE_{e^+}$ is the
positron energy spectrum per decay of the DM. 
The function $f_{e^+} (E)$ is now given by
\begin{eqnarray}
  f_{e^+} (E) &=&  \frac{1}{b(E)} \, \int_E^{E_{\rm max} }\; 
d E' \, I(\lambda_{D}(E,E') )\, Q_{\rm dec} (E') \nonumber \\
 &=& \frac{1}{b(E)} \, 
 \frac{1}{\tau_{\rm dm} } \,\left(  \frac{\rho_{\rm dm} }{M_{\rm dm}} \right ) 
 \, \int_E^{E_{\rm max} }\;   d E' \, I(\lambda_{D}(E,E') )\, \frac{dN_{e^+}}
{ d E'_{e^+}} \;,
\end{eqnarray}
with $E_{\rm max} = M_{\rm dm} / 2$ for the decaying dark matter and
summation over all decay channels is explicit.
The flux is the same as in Eq.(\ref{semiflux}). 
%

In our analysis, we use either a (i) monochromatic decaying spectrum:
\begin{equation}
  \frac{d N_{e^+} }{d E_{e^+}} (E) = \frac{2}{M_{\rm dm}} 
\delta \left ( 1 - \frac{2 E}{M_{\rm dm} }
\right ) \;,
\end{equation}
or (ii) varying decaying spectrum 
\begin{equation}
  \frac{d N_{e^+} }{d E_{e^+}} (E) = \frac{80 E }{ M^2_{\rm dm} }\,
   \left ( 1 - \frac{2 E}{M_{\rm dm} } \right )^3  \;.
\end{equation}
The exact form of the varying decaying spectrum is not crucial in our 
analysis.  As long as it is soft it suffices to suit our purpose.
We will show that the model that we will use in this work gives 
an energy spectrum consistent with the above varying spectrum.
Analogous formulas for the electron 
will be omitted as before.

\subsection{Background fluxes}

The background electron/positron fluxes from astrophysical sources are 
believed to be mainly due to
Supernova explosions for the primary electrons and 
from the interactions between the cosmic ray nuclei (mainly proton and helium) 
and atoms (mainly hydrogen and helium) in the interstellar medium 
for the secondary electrons and positrons. 
They are commonly parametrized as \cite{bkgflux}
\begin{eqnarray}
\Phi_{e^+}^{\rm bkdg} &=& \frac{ 4.5 E^{0.7} }{ 1 + 650 E^{2.3} + 1500 E^{4.2} }
 \; \; ,
\\
\Phi_{e^-}^{\rm bkdg} &=&  \Phi_{e^-}^{\rm bkdg, prim} + \Phi_{e^-}^{\rm bkdg, sec}
 \nonumber \\
&=& 
\frac{ 0.16 E^{-1.1} }{ 1 + 11 E^{0.9} + 3.2 E^{2.15} } +
\frac{ 0.7 E^{0.7} }{ 1 + 110 E^{1.5} + 580 E^{4.2} }  \;,
\end{eqnarray}
where $E$ is in unit of GeV and the unit for the flux is 
GeV$^{-1}$ cm$^{-2}$ s$^{-1}$
sr$^{-1}$.  We use a normalization of $0.7$ so that the
flux calculation is consistent with the ATIC data in the low energy 
range of $20-70$ GeV.
\footnote{ The normalization of background can vary
  between 0.6 and 0.8 so that our conclusion of 1-peak-1-bump
  structure is still valid. If we also vary the normalization in the
  fits, we found that the best fit for the ATIC data (M2) is similar
  to the result in row \#4 in Table II and the normalization is
  $0.69$, which is close enough to our fixed value of $0.7$.}

\subsection{Propagation models}

To evaluate the halo function $I(\lambda_D)$, 
we will use the Navarro-Frenk-White (NFW) halo profile with the 
propagation models ``M2'',
''Med'' and ``M1'' as parametrized in Ref. \cite{cirelli}.  The models ``M2'',
``Med'' and ``M1'' are characterized by the propagation length in the
increasing order. 
The analysis we will perform in the next section can be 
straightforwardly repeated for the other popular halo models 
specified by the Moore profile and the cored isothermal profile.

\section{Analysis}

\begin{figure}[th!]
\centering
\includegraphics[width=5in]{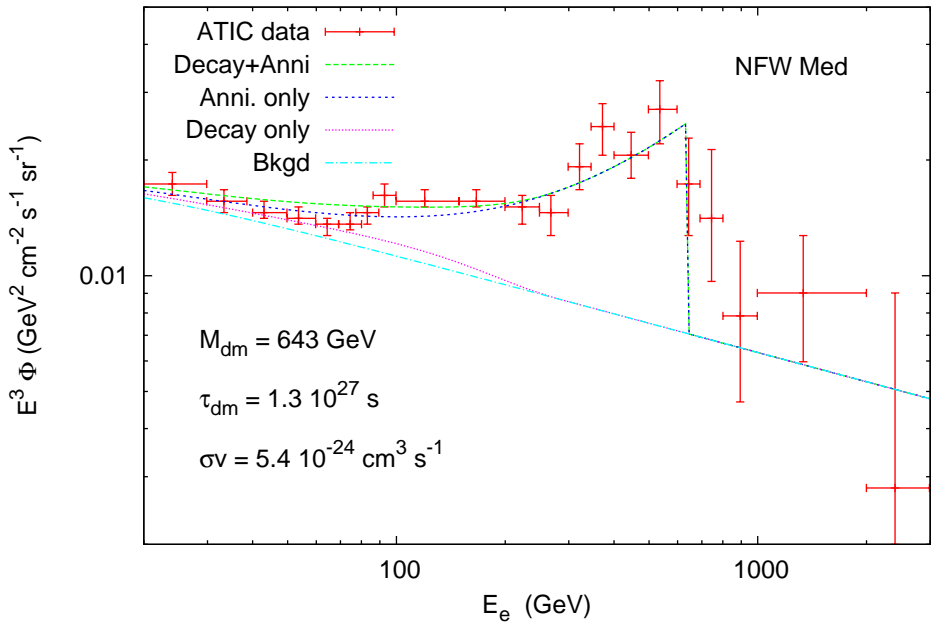}
\caption{\small \label{fig1} The spectrum for the ATIC data.  The dark matter
mass is taken to be 643 GeV, with a monochromatic spectrum for 
${\rm DM}\,  {\rm DM} \to e^+ e^-$ annihilation and a soft decaying spectrum of
$dN/dE \sim E ( 1 -  2E/M_{\rm dm})^3 /M_{\rm dm}^2$ for
${\rm DM} \to e^+ e^- X $ decay.}
\end{figure}

\begin{figure}[th!]
\centering
\includegraphics[width=5in]{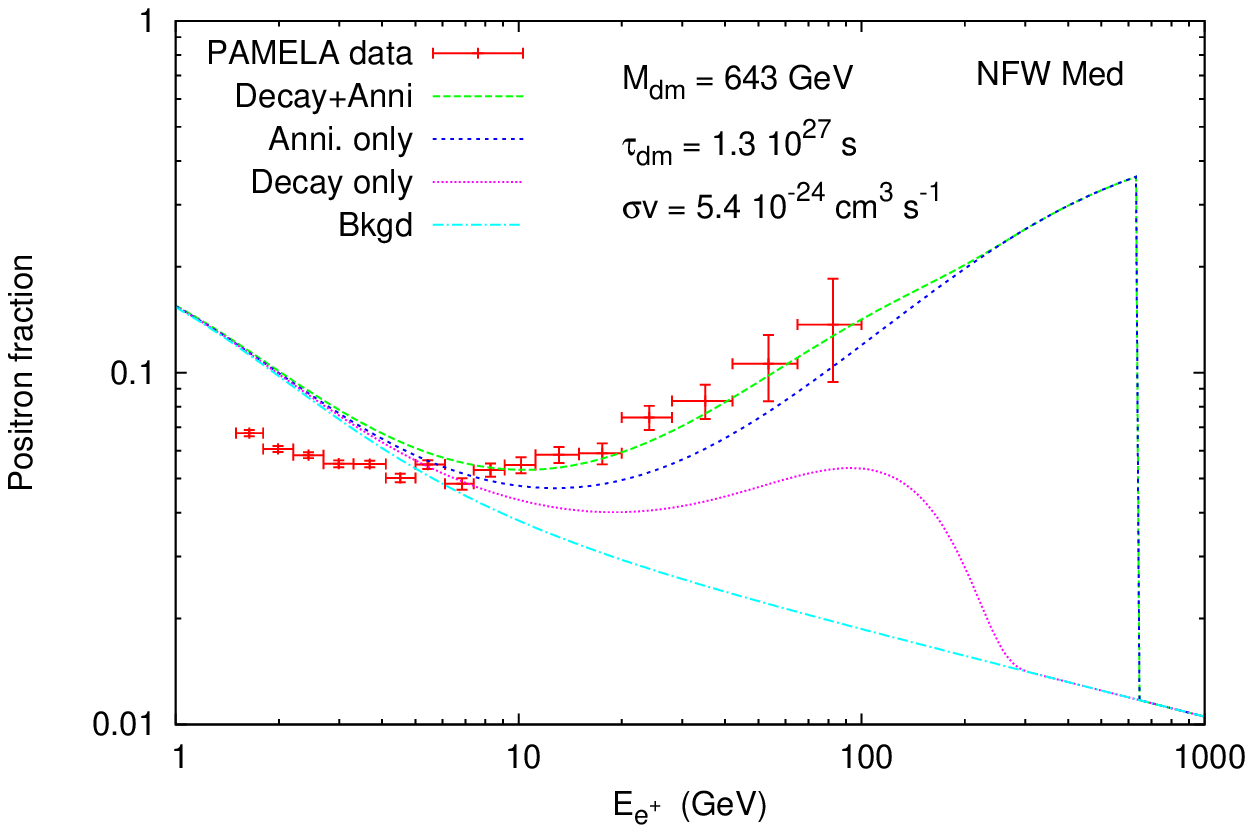}
\caption{\small \label{fig2} 
The positron fraction for the PAMELA data.   The dark matter
mass is taken to be 643 GeV, with a monochromatic spectrum for 
${\rm DM}\,  {\rm DM} \to e^+ e^-$ annihilation and a soft decaying spectrum of
$dN/dE \sim E ( 1 -  2 E/M_{\rm dm})^3 /M_{\rm dm}^2$ for
${\rm DM} \to e^+ e^- X $ decay.}
\end{figure}

Theoretical predictions of the energy spectrum and the positron fraction 
depend in general on the three input parameters 
$\tau_{\rm dm}$, $\langle \sigma v \rangle$ and $M_{\rm dm}$. 
To achieve the output more or less consistent with the ATIC and PAMELA 
data, we use $\tau_{\rm dm} =1.3 \times 10^{27}$ s,
$B\langle \sigma v \rangle = 5.4 \times 10^{-24}\;{\rm cm}^3 \, {\rm s}^{-1}$
and $M_{\rm dm}$ = 643 GeV.  The energy spectrum for ATIC 
and the positron fraction for PAMELA are  shown in the 
Fig. \ref{fig1} and Fig. \ref{fig2} respectively.  
We have used
the NFW halo model with propagation model ``Med'' in the figures.  From 
these two plots it is interesting to see that both PAMELA and ATIC data can be 
fitted simultaneously by the double-action dark matter. 
We now want to justify this fact more quantitatively using the technique of 
$\chi^2$ fits.

\subsection{Fitting with PAMELA data only}

The PAMELA data shown in Fig. \ref{fig2} has a rising trend
starting from point \#9 to \#16 which we will be using 
exclusively in our numerical analysis.  
The first eight data points are of low energy less than about 10 GeV,
where the solar activity of magnetic polarity state
is expected to play a significant role in the positron 
abundance \cite{pamela-e}.
Both horizontal and vertical errors are 
explicitly given in the PAMELA publication \cite{pamela-e} but only 
the latter one will be taken into account in our analysis.
The parameters used in the analysis are  
$\tau_{\rm dm}$, $\langle \sigma v \rangle$,
and $M_{\rm dm}$ as mentioned earlier.  
As indicated in previous section, 
we use a monochromatic electron/positron spectrum 
for dark matter annihilation, while for 
decaying dark matter we use either a
(i) monochromatic (mono) or (ii) varying spectrum (var).  It is
denoted by ``mono'' or ``var'' in the tables.
However, we emphasis that the results do not depend strongly 
on the exact spectrum as long as it
is soft enough.  We will show that in some cases the soft spectrum actually
fits better than the monochromatic spectrum.

In Table \ref{table-pam}, we show the fits to the
PAMELA data (point \#9 to \#16) with one or two of the parameters fixed.
It can be seen that the ``M1'' propagation model fits slightly better than 
``Med'' but much better than ``M2''.  When we vary the mass of dark matter
from 200 GeV to 1000 GeV, the goodness of the fit, measured by the $\chi^2$,
is roughly independent of the mass. (It shows a slight better fit when
$M_{\rm dm}$ increases, but not of any significance.)

\begin{table}[bh!]
\caption{\small Fitting with PAMELA data only. \label{table-pam}
When we fix $\tau_{\rm dm} = 10^{40}$ s, the contribution from decaying
is negligible.  Similarly, when we fix $B\langle \sigma v \rangle = 10^{-40}
\; {\rm cm}^3 \; {\rm s}^{-1}$ the contribution from annihilation is
negligible.  The ``mono'' means monochromatic spectrum,
whereas ``var'' means varying spectrum for the decaying contribution.
}
\begin{ruledtabular}
\begin{tabular}{l|lll|ll}
Prop. & $\tau_{\rm dm}$  &  $B\langle \sigma v \rangle $  &  $ M_{\rm dm}$
            &  $\chi^2$/\# d.o.f. & Comments \\
model &  (s)            & (cm$^{3}$ s$^{-1}$)  & (GeV)  & & \\
\hline
            & $10^{40}$ (fixed)  & $10^{-23}$  & 300 (fixed) & 79.6/7 & mono \\
NFW M2      & $0.45\cdot10^{27}$ & $10^{-40}$ (fixed) & 212 & 18.1/6  &  var \\
            & $0.55\cdot10^{27}$ & $0.77\cdot10^{-24}$ & 200 (fixed) & 10.1/6
 & var\\
\hline
            &   & $0.15\cdot10^{-23}$   & 250 (fixed) & 3.0/7 & mono\\
            & fixed   & $0.21\cdot10^{-23}$   & 300 (fixed) & 2.9/7 & mono\\
NFW Med     & at  & $0.37\cdot10^{-23}$   & 400 (fixed) & 2.8/7 & mono\\
            & $10^{40}$ & $0.83\cdot10^{-23}$   & 600 (fixed) & 2.7/7 & mono\\
            &    & $0.15\cdot10^{-22}$   & 800 (fixed) & 2.7/7 & mono\\
            &    & $0.22\cdot10^{-22}$   & 1000 (fixed) & 2.6/7 & mono\\
          \hline 
            &  fixed  &$0.59\cdot10^{-24}$   & 200 (fixed) & 3.6/7 & mono\\
NFW M1      &  at     &$0.36\cdot10^{-23}$   & 500 (fixed) & 2.2/7 & mono\\
            &  $10^{40}$ &$0.70\cdot10^{-23}$   & 700 (fixed) & 2.0/7 & mono\\
            &    &$0.14\cdot10^{-22}$   & 1000 (fixed) & 1.9/7 & mono
\end{tabular}
\end{ruledtabular}
\end{table}

Since the PAMELA data did not show any peak structure, the 
data would not
prefer any mass of the dark matter.  As long as the propagation diffuses
the spectrum or the spectrum itself is soft, it can fit the data well. 
It is shown in the first three rows in Table \ref{table-pam} when we used
the varying spectrum.  The ``var'' spectrum fits much better than the ``mono''
spectrum.  
Thus, the PAMELA data alone do not constrain the mass of dark matter to any
significant extent according to this analysis, 
as long as the mass is heavier than about 200 GeV.

\subsection{Fitting with ATIC data only}

The ATIC data exhibits a more interesting feature of one peak plus one 
hump structure.  Thus, we expect
that the data prefer some mass range of the dark matter for both
annihilation and decaying contributions.  
By comparing row \#4 with \#1 to \#3 in Table \ref{table-atic} where 
the ``M2'' propagation model is used, 
we note that with a dark matter about 744 GeV 
the fit is substantially better with both annihilation and soft 
decaying contributions included than just either one of them is used.  
On the other hand, the other two propagation models do not show such effects.
The less diffuse propagation model ``M2'' fits better than the ``Med'' and
``M1'' models.

\begin{table}[bh!]
\caption{\small Fitting with ATIC data only. Other details are the same
as Table \ref{table-pam}.
\label{table-atic}}
\begin{ruledtabular}
\begin{tabular}{l|lll|ll}
Prop. & $\tau_{\rm dm}$  &  $B\langle \sigma v \rangle $  &  $ M_{\rm dm}$
            &  $\chi^2$/\# d.o.f. & Comments \\
model &  (s)            & (cm$^{3}$ s$^{-1}$)  & (GeV)  & & \\
\hline
        & $10^{40}$ (fixed)  & $0.51\cdot10^{-23}$  & 536 & 39.5/19 & mono \\
NFW M2  & $0.36\cdot10^{27}$ & $10^{-40}$ (fixed) & 1072 & 39.5/19  & mono \\
        & $1.0\cdot10^{26}$ & $10^{-40}$ (fixed) & 3190 & 25.3/19  & var \\
        & $0.65\cdot10^{27}$ & $0.80\cdot10^{-23}$ & 744 & 23.7/18  & var \\
\hline
    & $10^{40}$ (fixed)  & $0.78\cdot10^{-23}$   & 745  & 27.4/19 & mono\\
NFW Med  &$0.32\cdot10^{27}$  & $10^{-40}$ (fixed) & 1490 & 27.4/19 & mono\\
         &$0.25\cdot10^{29}$  & $0.78\cdot10^{-23}$& 745  & 27.4/18 & var\\
 \hline 
         &$10^{40}$ (fixed)  &$0.68\cdot10^{-23}$ & 740 & 34.2/19 & mono\\
NFW M1   &$0.36\cdot10^{27}$ &$10^{-40}$ (fixed) & 1470 & 34.2/19 & mono\\
         &$0.11\cdot10^{27}$ &$10^{-40}$ (fixed) & 4420 & 37.6/19 & var
\end{tabular}
\end{ruledtabular}
\end{table}

\subsection{Fitting using both PAMELA and ATIC data}

The fits are shown in Table \ref{table-all}.  
\begin{table}[bh!]
\caption{\small Fitting with PAMELA and ATIC data. Other details are the same
as previous tables. \label{table-all}}
\begin{ruledtabular}
\begin{tabular}{l|lll|ll}
Prop. & $\tau_{\rm dm}$  &  $B\langle \sigma v \rangle $  &  $ M_{\rm dm}$
            &  $\chi^2$/\# d.o.f. & Comments \\
model &  (s)            & (cm$^{3}$ s$^{-1}$)  & (GeV)  & & \\
\hline
  & $10^{40}$ (fixed)  & $0.31\cdot10^{-23}$  & 400 (fixed) & 297/28 & mono \\
  &$10^{40}$ (fixed) & $0.47\cdot10^{-23}$ & 500 (fixed) & 281/28 & mono\\
    &$10^{40}$ (fixed) & $0.70\cdot10^{-23}$  & 600 (fixed) & 269/28  & mono \\
NFW M2 &$10^{40}$ (fixed) & $0.94\cdot10^{-23}$ & 700 (fixed)& 275/28  & mono \\
    &$10^{40}$ (fixed) & $0.12\cdot10^{-22}$ & 800 (fixed)& 284/28  & mono \\
    &$10^{40}$ (fixed) & $0.16\cdot10^{-22}$ & 1000 (fixed) & 343/28  & mono \\
    &$10^{40}$ (fixed) & $0.55\cdot10^{-23}$ & 535  & 267/27  & mono \\
\hline
    & $10^{40}$ (fixed)  & $0.27\cdot10^{-23}$   & 400 (fixed)& 96.8/28 & mono\\
   &$10^{40}$ (fixed) &$0.42\cdot10^{-23}$  & 500 (fixed)& 74.3/28 & mono\\
      &$10^{40}$ (fixed) &$0.60\cdot10^{-23}$  & 600 (fixed)& 59.1/28 & mono\\
      &$10^{40}$ (fixed) &$0.80\cdot10^{-23}$  & 700 (fixed)& 59.3/28 & mono\\
NFW Med &$10^{40}$ (fixed) &$0.10\cdot10^{-22}$  & 800 (fixed)& 63.2/28 & mono\\
      &$10^{40}$ (fixed) &$0.15\cdot10^{-22}$  & 1000 (fixed)& 102/28 & mono\\
    &$0.16\cdot10^{28}$ &$0.62\cdot10^{-23}$  & 745 & 55.9/26 & mono\\
    &$0.13\cdot10^{28}$ &$0.54\cdot10^{-23}$  & 643 & 41.9/26 & var\\
 \hline 
     &$10^{40}$ (fixed)  &$0.34\cdot10^{-23}$ & 500 (fixed) & 55.3/28 & mono\\
   &$10^{40}$ (fixed) &$0.48\cdot10^{-23}$ &600 (fixed) & 41.1/28 & mono\\
       &$10^{40}$ (fixed) &$0.64\cdot10^{-23}$ &700 (fixed) & 38.8/28 & mono\\
       &$10^{40}$ (fixed) &$0.83\cdot10^{-23}$ &800 (fixed) & 40.0/28 & mono\\
NFW M1  &$10^{40}$ (fixed) &$0.99\cdot10^{-23}$ &900 (fixed) & 66.9/28 & mono\\
       &$10^{40}$ (fixed) &$0.12\cdot10^{-22}$ &1000 (fixed) &66.7/28 & mono\\
       &$0.38\cdot10^{28}$&$0.61\cdot10^{-23}$ & 745  &38.5/26 & mono\\
       &$0.33\cdot10^{29}$&$0.54\cdot10^{-23}$ & 643  &38.6/26 & var\\
\end{tabular}
\end{ruledtabular}
\end{table}
Similar to the fits with
PAMELA data only, the fits using ``M1'' propagation model are 
slightly better than using
``Med'', which are in turns much better than ``M2''.
The fits using ``M2'' is not good at all, given the fact that 
$\chi^2$ per d.o.f. is large.   This behavior is similar to those fitted
to PAMELA data alone.

In the fits with ``Med'' propagation model, we first fitted with a 
negligible decaying contribution, i.e., with annihilation contribution 
only.  The best that we can do is $\chi^2\simeq 59$ at 
$M_{\rm dm} \simeq 600-700 $ GeV.  When we also 
turn on the decaying contribution with a varying spectrum, the $\chi^2$
goes down to $42$.
This result supports a dark matter that annihilates 
and also decays with a soft varying spectrum.  
The best fit is 
\begin{equation}
\tau_{\rm dm} = 0.13\cdot 10^{28}\; {\rm s}, \qquad
B\langle \sigma v \rangle = 0.54 \cdot 10^{-23}\; {\rm cm}^3\; {\rm s}^{-1},
\qquad
M_{\rm dm} = 643\; {\rm GeV}
\end{equation}
which has a $\chi^2 \simeq 42/26$ d.o.f.  
We used this set of fitted parameters in Figs. \ref{fig1}
and \ref{fig2}.

The fits using the ``M1'' propagation model improve further. 
Features are similar to the case of ``Med''.  
The best fit using monochromatic annihilation and decaying contributions is 
(second last row of Table \ref{table-all})
\begin{equation}
\tau_{\rm dm} = 0.38\cdot 10^{28}\; {\rm s}, \qquad
B\langle \sigma v \rangle = 0.61 \cdot 10^{-23}\; {\rm cm}^3\; {\rm s}^{-1},
\qquad
M_{\rm dm} = 745\; {\rm GeV}
\end{equation}
with a $\chi^2= 38.5/26$ d.o.f.
The best fit using monochromatic annihilation but varying spectrum for 
decaying contribution is (last row of Table \ref{table-all})
\begin{equation}
\tau_{\rm dm} = 0.33\cdot 10^{29}\; {\rm s}, \qquad
B\langle \sigma v \rangle = 0.54 \cdot 10^{-23}\; {\rm cm}^3\; {\rm s}^{-1},
\qquad
M_{\rm dm} = 643\; {\rm GeV}
\end{equation}
with a $\chi^2= 38.6/26$ d.o.f.

\section{A model}
\begin{figure}[th!]
\centering
\includegraphics[width=3in]{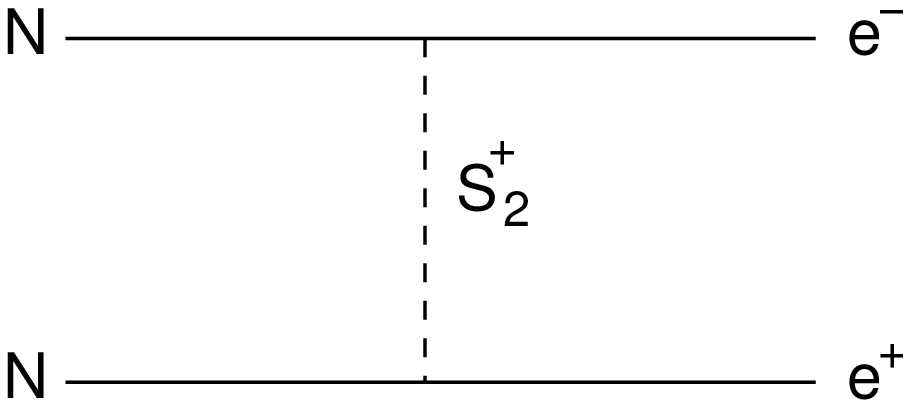}
\includegraphics[width=3in]{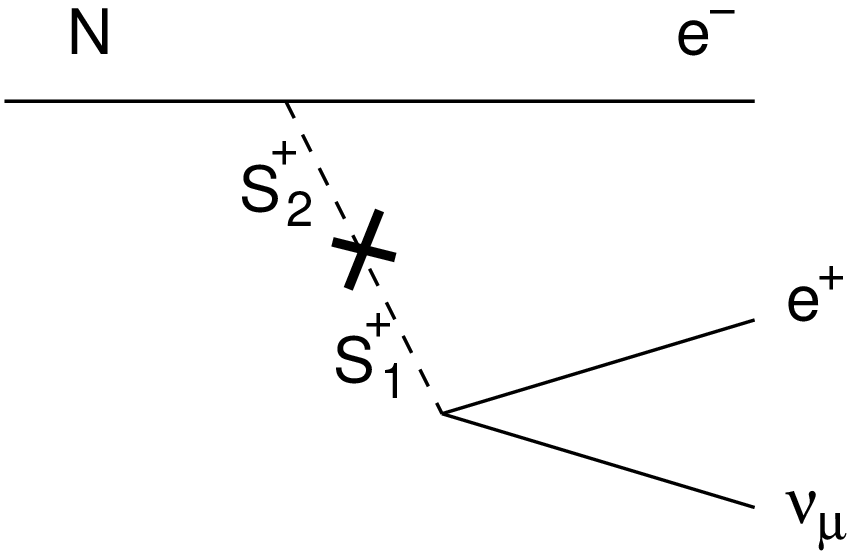}

(a)  \hspace{3in}  (b)
\caption{\small \label{fey} Feynman diagrams for 
(a) annihilation $N N \to e^- e^+$ and (b) decay $N \to e^- e^+ \nu_\mu$.
}
\end{figure}

\begin{figure}[th!]
\centering
\includegraphics[width=5in,clip]{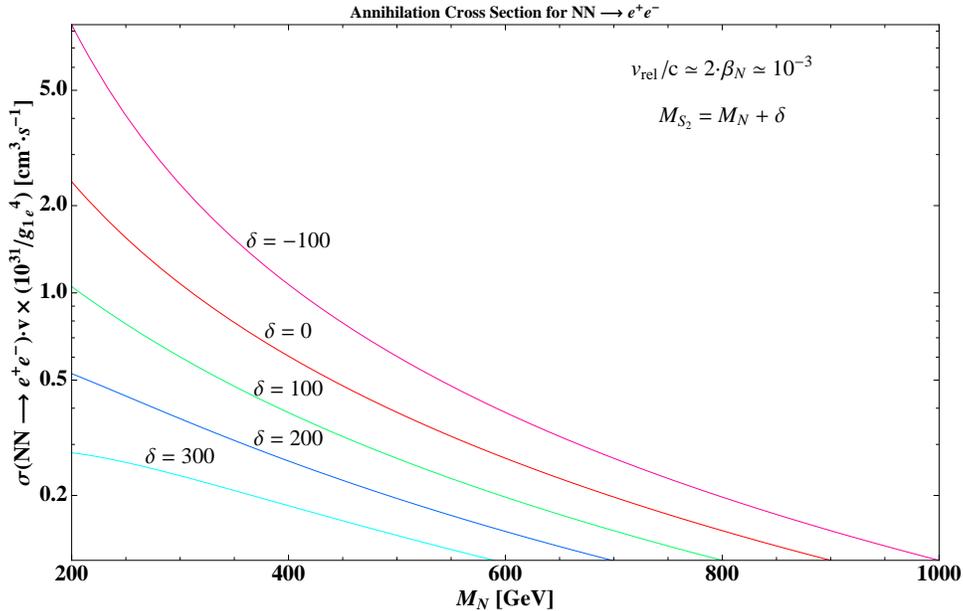}
\caption{\small \label{newfig}
Annihilation cross section versus 
the right-handed neutrino mass for various values of the scalar mass
at a center-of-mass energy $\sqrt{s}$ when the $v_{\rm rel} \approx 10^{-3}$.}
\end{figure}

We use a TeV right-handed neutrino mass model to present a dark matter
candidate that can annihilate and decay.  The model can be described by
the following interaction Lagrangian \cite{Krauss,seto}
\begin{eqnarray}
{\cal L}_{\rm int} &=& f_{\alpha\beta} L^T_\alpha C i \tau_2 L_\beta S^+_1
   + g_{1\alpha} N_1 S_2^+ \ell_{\alpha R}
   + g_{2\alpha} N_2 S_2^+ \ell_{\alpha R}  + {\rm H.c.} \nonumber \\
&& + \, M_{N_1} N^T_1 C N_1 + M_{N_2} N^T_2 C N_2 - V(S_1,S_2)
\label{lag}
\end{eqnarray}
where $L_{\alpha,\beta}$ and $l_{\alpha R}$ are the lepton doublet and
singlet respectively with $\alpha,\beta$ denoting the family indices,
$N_{1,2}$ are the two right-handed neutrinos, $C$ is the
charge-conjugation operator, and $V(S_1,S_2)$ is the scalar potential
for the two complex scalar fields $S_1$ and $S_2$ that containing a
term $\lambda_s (S_1^+ S_2^-)^2$.  Note that $f_{\alpha\beta}$ is
antisymmetric under the interchange of the family indices.  The original
model in Ref. \cite{Krauss} contains only one $N_1$, but the improvement
in Ref. \cite{seto} by adding another $N_2$ makes the model consistent
with the neutrino oscillation data.  For simplicity we shall only
describe the lighter right-handed neutrino $N_1$ (denoted simply by $N$ in 
what follows), which is
in the TeV mass range, as far as dark matter is concerned.  It
has been shown \cite{seto} that $N$ can be a dark
matter candidate.  The annihilation $N N \to e^+ e^-$ can go through $t$- and
$u$-channel diagrams with an intermediate $S_2^+$ (see
Fig. \ref{fey}(a)) with the rate given by
\begin{eqnarray} 
\sigma v_{\rm rel} &=& \frac{g_{1e}^4 }{64\pi}\, \frac{1}{s} \; 
\int_{-1}^{1} \; d x \; \Biggr \{
 \frac{ s^2 ( 1- \beta_N x)^2}{ 4 ( M_N^2 - M_{S_2}^2 - \frac{s}{2}
 ( 1- \beta_N x) )^2 } + 
 \frac{ s^2 ( 1+  \beta_N x)^2}{ 4 ( M_N^2 - M_{S_2}^2 - \frac{s}{2}
 ( 1 + \beta_N x) )^2 } \nonumber \\
&-& \frac{2 M_N^2 s}{( M_N^2 - M_{S_2}^2 - \frac{s}{2} ( 1- \beta_N x) )
( M_N^2 - M_{S_2}^2 - \frac{s}{2} ( 1 + \beta_N x) ) }  \Biggr \}
\end{eqnarray}
where $\beta_N = \left( 1 -  4 M^2_N / s \right)^{1/2}$. 
As $\beta_N \to 0$, the above annihilation rate vanishes.
This is expected for the annihilation rate for identical Majorana fermions 
is P-wave suppressed \cite{haim}.
When the center of mass energy $\sqrt{s}$ is slightly above the 
threshold of $2 M_N$, the 
annihilation electron/positron energy spectrum is almost a monochromatic one.
We show in Fig. \ref{newfig} the annihilation cross section versus 
the right-handed neutrino mass for various values of the scalar mass,
taking the center-of-mass energy $\sqrt{s}$ at which
$v_{\rm rel} \approx 10^{-3}$.
The plot demonstrates the P-wave suppression. 
Therefore, a large boost factor of order $O(10^7)$
is needed to fit the data in this model.
However, this suppression might not be taken literally.
If one is willing to extend the model by introducing new long range
force among the dark matter, the Sommerfeld enhancement for P-wave
annihilation can be significantly larger than the S-wave case \cite{pwave}.

\begin{figure}[th!]
\centering
\includegraphics[width=5in]{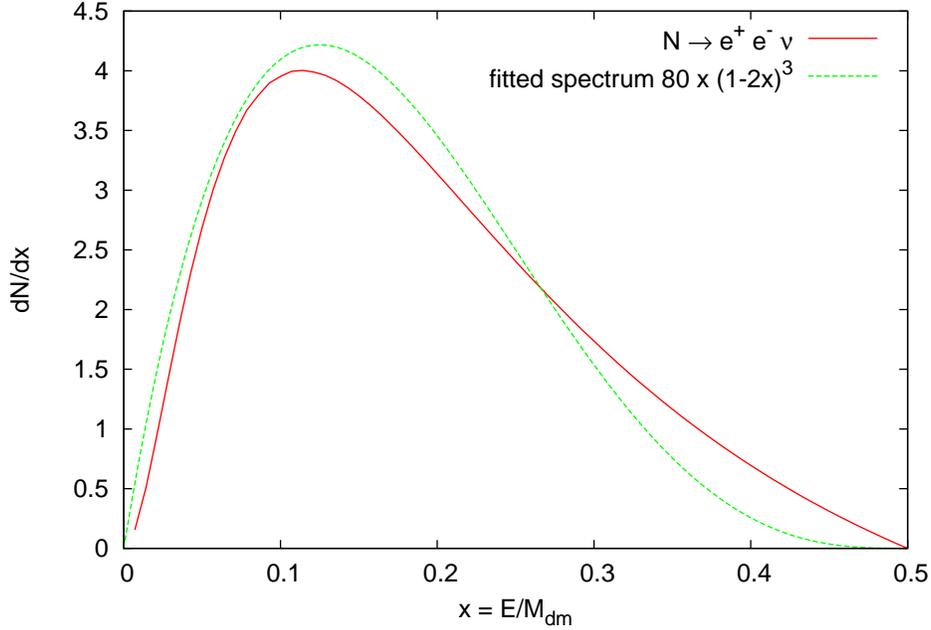}
\caption{\label{Ndecay}
Normalized energy spectrum $1/\Gamma d\Gamma/dx$ with $x=E/M_{\rm dm}$ 
for the decay $N \to e^- e^+ \nu_\mu$ and the approximation 
$80 x ( 1 - 2x)^3$.}
\end{figure}

In Refs. \cite{Krauss,seto}, the $N$ is assumed stable by imposing a $Z_2$
parity.  Here we introduce a small violation of this parity by adding a term
$\epsilon S^+_1 S^-_2 + {\rm H.c.}$ to the scalar potential, where 
$\epsilon \sim (1\;{\rm eV})^2$, which is of the order of the square of 
neutrino mass.  The decay of $N$ can then go through a Feynman diagram
shown in Fig. \ref{fey}(b).
\footnote{
There is another decay channel $N\to e^- \mu^+ \nu_e$ given the same
couplings. The lifetime will be shorten by a factor of two, but it
does not affect our order-of-magnitude estimate.
}
An order of magnitude estimate of the decay width of the $N$ can be given as
\begin{equation}
 \Gamma_{N} \sim g_{1e}^2 f_{12}^2 \epsilon^2 M_N / M_{S_2}^4  \;.
\end{equation}
Taking typical values of the couplings \cite{seto} 
($g_{1e} \sim 10^{-1},\; f_{12} \sim 10^{-2},\; M_{N} \sim M_{S_2} \sim$ TeV), 
the lifetime of $N$ is roughly
\begin{equation}
\tau_{N} \sim 10^{26}\; \left ( \frac{ {\rm eV}^2} {\epsilon} \right )^2
   \; {\rm sec} \;.
\end{equation} 
It is interesting to see that when $\epsilon$ is of the order of the
neutrino mass squared, the amount of violation of the $Z_2$ parity
is in the right order to fit the data.
We calculate the normalized energy spectrum of the decay
$N \to e^- e^+ \nu_\mu$ shown in Fig. \ref{Ndecay}, where the
approximate spectrum $80 x ( 1-2x)^3$ with $x = E/M_{\rm dm}$ is also
shown.  The figure justifies the approximation of the energy spectrum
that we have used in our analysis given in the previous section.  
Note that the exact form is not 
crucial in the fits as long as the spectrum is soft.
In this model, the decay and annihilation of the dark matter 
are pure leptonic.  It will not give enhancement to the $\bar p$ flux.

\section{Conclusion}

We have pointed out a 2-bump (or 1 peak plus 1 hump) structure in the
ATIC data.  If this feature can be sustained it cannot be explained by 
dark matter 
annihilation or decay alone.  It can be either a two-component dark
matter or the dark matter can undergo both annihilation and decay at the
same time.  We have shown that such a double-action dark matter can fit better
to the ATIC and PAMELA data simultaneously than just annihilation or decay
alone.  We have employed a TeV right-handed neutrino model to illustrate
the idea. The original model only allows dark matter annihilation, but
here we have introduced a small breaking of the $Z_2$ parity at the order
of the neutrino mass squared.  With such a coincident size of $Z_2$ breaking
one can explain the long lifetime required to fit the data.
This indicates there may be intricate connection between neutrino mass 
problem with dark matter physics. We close with a few comments.

\begin{itemize}

\item 
The model of TeV right-handed neutrino 
can account for the neutrino mass and oscillation pattern
\cite{Krauss,seto}.  It is also consistent with lepton-flavor violation.
The lightest of the right-handed neutrino can be a dark matter candidate
and its relic density can account for the observed dark matter density.

\item
Reference \cite{seto} showed that the model can be made consistent 
with existing
neutrino oscillation data by tuning the parameters of the model.  If
so there would also be the muon and tau channels in the annihilation.
The electron/positron spectrum would contain the monochromatic part
and the continuous part from muon and tau decays. The $\chi^2$ analysis
would be much more involved and beyond the scope of this paper.

\item 
The decay and annihilation of the dark matter 
are pure leptonic.  It will not give enhancement to the $\bar p$ flux.

\item The $Z_2$ parity violation parameter $\epsilon$ that we introduced is,
by accident, at the order of the square of the neutrino mass.

\item The prediction for the gamma-ray flux mainly comes from the 
bremsstrahlung off the electron/positron.

\item Since the TeV right-handed neutrino has no appreciable coupling
to quarks or gluons, the scattering cross section with nuclei is negligible.
Thus, the sun or the Earth will not capture any large amount of dark matter,
and so no enhancement to the neutrino flux coming from the core of the sun.

\item Either annihilation or decay alone cannot explain the probable
1-peak plus 1-hump structure in the ATIC data.  We have shown that 
a dark matter that can annihilate and decay simultaneously can explain
the ATIC and PAMELA data at the same time.

\end{itemize}

Dark matter interpretation for the PAMELA and ATIC experiments is
exciting since it implies new physics beyond the Standard Model.
However, one should keep in mind that a yet unidentified astrophysical
object such as a nearby pulsar or micro-quasar could be a primary
source as well.  Furthermore, many models are capable to explain the
excess anomaly.  PAMELA is extending the spectra measurement to higher
energy of about 300 GeV for positrons and 500 GeV for electrons.  New
data is also expected soon from the FERMI satellite for the diffuse
Galactic cosmic $\gamma$-ray spectrum.  These future developments will
certainly help to discriminate models and unravel the true nature of
the anomaly seen thus far. More excitements are waiting ahead of us!

\section*{Acknowledgment}
We thank Wai-Yee Keung and Vernon Barger for a useful communication.
The work was supported in parts by the NSC of Taiwan under grant
no. 96-2628-M-007-002-MY3, the NCTS, the Boost Program of NTHU, and
the WCU program through the KOSEF funded by the MEST
(R31-2008-000-10057-0).

\end{document}